\documentclass[aps,prl,twocolumn,groupedaddress]{revtex4}

\usepackage{graphicx}
\usepackage{dcolumn}
\usepackage{bm}

\begin{document}

\preprint{Cabello 2colonne vere.tex}

\title{{\bf Experimental evidence for bounds on quantum correlations}}

\author{  F. A. Bovino }
  \email{fabio.bovino@elsag.it}
\author{ G. Castagnoli }

\affiliation{
Elsag spa \\
Via Puccini 2-16154 Genova (Italy) }

\author{ I. P. Degiovanni }
  \email{ degio@ien.it }
\author{ S. Castelletto }


\affiliation{
Istituto Elettrotecnico Nazionale G. Ferraris \\
Strada delle Cacce 91-10135 Torino (Italy) }

\date{\today}

\begin{abstract}

We implemented the experiment proposed by Cabello
[arXiv:quant-ph/0309172] to test the bounds of quantum
correlation. As expected from the theory we found that, for
certain choices of local observables, Cirel'son's bound of the
Clauser-Horne-Shimony-Holt inequality ($2\sqrt{2}$) is not reached
by any quantum states.

\end{abstract}


\maketitle

The philosophical debate on the completeness of quantum mechanics,
on the existence of local-realistic hidden variable theories was
originated by the famous EPR paper \cite{epr}. The Bohm's version
of EPR argument \cite{bohm} dealing with the quantum correlation
of two-particle singlet state, triggered Bell's derivation of an
experimentally testable inequality, in principle allowing the
discrimination between the quantum world and the classical
local-realistic one \cite{bell}.

Since then, several Bell's inequalities using two or more
particles have been proposed \cite{CHSH,otine}, and a lot of
experiments with different quantum systems have been performed
showing violation of Bell's inequality in good agreement with the
predictions of quantum mechanics \cite{bellexp}.

One of the most widely known Bell's inequalities, the
Clauser-Horne-Shimony-Holt (CHSH) inequality \cite{CHSH}, states
that, for two separate particles $a$ and $b$, a necessary and
sufficient condition to describe the correlation between the two
particles by a local-realistic theory is
\begin{equation}
|S|=|\langle \widehat{O}^{(1)}_{a} \widehat{O}^{(I)}_{b} \rangle +
\langle \widehat{O}^{(2)}_{a} \widehat{O}^{(I)}_{b} \rangle +
\langle \widehat{O}^{(1)}_{a} \widehat{O}^{(II)}_{b} \rangle -
\langle \widehat{O}^{(2)}_{a} \widehat{O}^{(II)}_{b} \rangle| \leq
2. \label{eq1
}
\end{equation}
$\widehat{O}^{(1)}_{a}$ and $\widehat{O}^{(2)}_{a}$
($\widehat{O}^{(I)}_{b}$ and $\widehat{O}^{(II)}_{b}$) are
physical observables corresponding to local measurement on
particle $a$ ($b$) with possible values of +1 or -1, and $\langle
\widehat{O}^{(i)}_{a} \widehat{O}^{(j)}_{b} \rangle$ is the
expectation value of the product of the two observables, i.e. it
accounts for the correlation between $\widehat{O}^{(i)}_{a}$ and
$\widehat{O}^{(j)}_{b}$.

Besides the limit imposed on $S$ by the classical correlation
described above, Cirel'son \cite{cirel} proved a further bound on
$S$ given by quantum correlation itself:
\begin{equation}
|S| \leq 2 \sqrt{2}.
\end{equation}
Cirel'son's inequality is only a necessary but not a sufficient
condition for correlations viable by quantum mechanics. This means
that for any quantum state of a two particles system quantum
mechanics predicts that the Cirel'son's inequality is satisfied,
but, for specific sets of observables, there are ranges of $S$
values, satisfying Cirel'son's inequality, which are not
attainable by any quantum system. In other words, the range of $S$
values corresponding to superquantum correlation -correlation
beyond the one predicted by quantum mechanics- are not restricted
to $2 \sqrt{2} < |S| \leq 4$, but, for specific choices of local
observables, region of superquantum correlation can be found also
within $2 < |S| \leq 2 \sqrt{2}$ \cite{cabellos}.

Although in Ref. \cite{masanes} the necessary and sufficient
condition for setting $S$ bounds according to quantum mechanics
was found, a practical characterization of these bounds was not
provided \cite{cabellos}. Filipp and Svozil \cite{filipp} proposed
a method for describing these bounds given a particular sets of
local observables, by using a computer simulation of a large
number of possible quantum systems chosen in a random way
\cite{filipp}. Cabello \cite{cabellos} found the analytical
formulation of the $S$ bounds according to the Filipp and Svozil
parameterization of the local observables in Eq. (\ref{eq1 })
given by
\begin{eqnarray}
\widehat{O}^{(1)}_{a} &=& \widehat{O}_{a} (2 \theta ) =  \cos(2 \theta)
\widehat{Z}_{a} + \sin(2 \theta) \widehat{X}_{a}  \nonumber \\
\widehat{O}^{(I)}_{b} &=& \widehat{O}_{b} ( \theta ) = \cos( \theta)
\widehat{Z}_{b} + \sin( \theta) \widehat{X}_{b}  \nonumber \\
\widehat{O}^{(2)}_{a} &=& \widehat{O}_{a} (0 ) = \widehat{Z}_{a}  \nonumber \\
\widehat{O}^{(II)}_{b} &=& \widehat{O}_{b} (3 \theta ) = \cos(3
\theta) \widehat{Z}_{b} + \sin(3 \theta) \widehat{X}_{b},
 \label{a1}
\end{eqnarray}
where $\widehat{X}$ and $\widehat{Z}$ are the usual Pauli's
matrices and $0 \leq \theta \leq \pi$. As quantum bounds can
always be attained using a suitably chosen entangled state,
Cabello in Ref. \cite{cabellos} shows that the bound of $S$,
according to Filipp and Svozil parameterization, is always reached
by a maximally entangled state belonging to the set
\begin{equation}
| \varphi (\xi) \rangle_{ab}= \cos(\xi) | \phi ^{+} \rangle_{ab} +
\sin(\xi) | \psi ^{-} \rangle_{ab}  \label{eq4}
\end{equation}
with $| \phi ^{+} \rangle_{ab}=2^{-1/2}(| H \rangle_{a} | H
\rangle_{b} + |V \rangle_{a} |  V \rangle_{b})$ and $| \psi ^{-}
\rangle_{ab}=2^{-1/2}(| H \rangle_{a} | V \rangle_{b} + |V
\rangle_{a} |  H \rangle_{b})$.

\begin{figure}[tbp]
\par
\begin{center}
\includegraphics[angle=0, width=7.5 cm, height=5 cm]{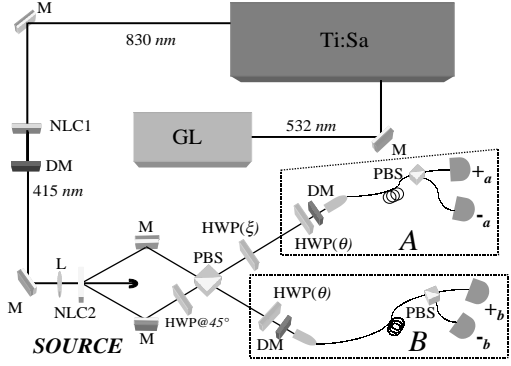}
\end{center}
\caption{ Experimental set-up: the time-compensated source of
photon pairs in the state $| \varphi (\xi) \rangle_{ab}$ is
composed of a type II nonlinear crystal (NLC2), pumped by the
pulsed laser system (LD, Ti:Sa and NLC1), of a polarizing beam
splitter (PBS), and of half-wave plates (HWP). Photons $a$ and $b$
of each pairs are directed towards the corresponding detection
apparatuses ("\textit{A}" and "\textit{B}") to perform the local
measurements, by means of HWPs, dichroic mirrors (DM) fiber
couplers, fibers integrated PBS, single-photon detectors. M
mirror, L lens} \label{Figure 1}
\end{figure}

Given this theoretical breakthrough, Cabello's proposal
\cite{cabellos} of an experiment to test the behavior of the
quantum correlations, is of straightforward interest, and in this
paper we perform this experiment by the setup depicted in Fig. 1.
In this scheme, the source of pulsed parametric down-conversion
(PDC) -to date the most easy method to generate two-particle
quantum correlation- is obtained by a 3 mm length BBO nonlinear
crystal (NLC2) with an average pumping power of 20 mW. Ultrashort
pump pulses (160 fs) at 415 nm are generated from the second
harmonic (NLC1) of a mode-locked Ti-Sapphire with repetition rate
of 76 MHz pumped by 532 nm green laser. PDC degenerate photon
pairs at 830 nm are generated by a non-collinear type II phase
matching for a 3.4$^{\circ}$ emission angle, providing eventually
a polarization entanglement, i.e. the singlet state $| \psi^{-}
\rangle_{ab}=2^{-1/2}(| H \rangle_{a} | V \rangle_{b} - |V
\rangle_{a} | H \rangle_{b}) $ \cite{kwiat2}, when
time-compensated PDC scheme is applied \cite{kimgrice}. To realize
the set of entangled states $| \varphi (\xi) \rangle_{ab}$ in Eq.
(\ref{fi}) an half-wave plate (HWP) on the channel $b$ is used to
rotate the polarization of photon $b$ of an angle $\xi- \pi/2$.

The local measurements on photon $a$ and $b$ are preformed by
identical apparatuses composed of open-air fiber couplers
collecting the PDC in single-mode optical fibers. HWPs before the
fiber coupler together with fiber-integrated polarizing beam
splitters (PBSs) project photons in the polarization basis $\{|s (
\alpha) \rangle = \cos(\alpha /2) |H \rangle + \sin(\alpha /2) |V
\rangle, |s^{\bot} ( \alpha) \rangle \ = \sin(\alpha /2) |H
\rangle - \cos(\alpha /2) |V \rangle \}$. Photons at the output
ports of the PBSs are detected by fiber coupled photon counters
(Perkin-Elmer SPCM-AQR-14) \cite{disc}. Dichroic mirror are placed
in front of the fiber couplers to reduce straylight, and
optimization of single-mode fiber collection, yielding the highest
final visibility in the experiment, is guaranteed by a proper
engineering of pump and collecting mode in the experimental
conditions \cite{fabiogianni}.

The Filipp and Svozil local observables can be rewritten for the
chosen polarization basis $\{|s ( \alpha) \rangle, |s^{\bot} (
\alpha) \rangle \}$ as
\begin{equation}
\widehat{O}(\alpha)= |s (\alpha) \rangle \langle s (\alpha) | -
|s^{\bot} (\alpha) \rangle \langle s^{\bot} (\alpha) |; \nonumber
\end{equation}
and thus, the correlation function $\langle
\widehat{O}_{a}(\alpha) \widehat{O}_{b}(\beta) \rangle$ in terms
of coincidence detection probabilities, $p_{x_{a}, y_{b}}(\alpha,
\beta, \xi)$ ($x,y=+,-$), as
\begin{widetext}
\begin{equation}
\langle \widehat{O}_{a}(\alpha) \widehat{O}_{b}(\beta) \rangle =
p_{+_{a}, +_{b}}(\alpha, \beta, \xi)
 + p_{-_{a}, -_{b}}(\alpha,
\beta, \xi) - p_{+_{a}, -_{b}}(\alpha, \beta, \xi)- p_{-_{a},
+_{b}}(\alpha, \beta, \xi) \nonumber
\end{equation}
\end{widetext}
where
\begin{eqnarray}
p_{+_{a}, +_{b}}(\alpha, \beta, \xi) &=& |\langle  \varphi (\xi)_{ab}
|s (\alpha) \rangle_{a} |s (\beta) \rangle_{b} |^{2},  \nonumber \\
p_{-_{a}, -_{b}}(\alpha, \beta, \xi) &=& |\langle  \varphi (\xi)_{ab}
|s^{\bot} (\alpha) \rangle_{a} |s^{\bot} (\beta) \rangle_{b} |^{2},  \nonumber \\
p_{+_{a}, -_{b}}(\alpha, \beta, \xi) &=& |\langle  \varphi (\xi)_{ab}
|s (\alpha) \rangle_{a} |s^{\bot} (\beta) \rangle_{b} |^{2},  \nonumber \\
p_{-_{a}, +_{b}}(\alpha, \beta, \xi) &=& |\langle  \varphi (\xi)_{ab}
|s^{\bot} (\alpha) \rangle_{a} |s (\beta) \rangle_{b} |^{2}.  \nonumber \\
\end{eqnarray}

$p_{x_{a}, y_{b}}(\alpha, \beta, \xi)$ are normalized in terms of
the number of coincident counts:
\begin{equation}
p_{x_{a}, y_{b}}(\alpha, \beta, \xi)=\frac{N_{x_{a}, y_{b}}(\alpha, \beta, \xi)}{%
\begin{array}{c}
\lbrack N_{+_{a}, +_{b}}(\alpha, \beta, \xi)+N_{+_{a}, -_{b}}(\alpha, \beta, \xi)+ \\
N_{-_{a}, +_{b}}(\alpha, \beta, \xi)+N_{-_{a}, -_{b}}(\alpha,
\beta, \xi)]
\end{array}
}
\end{equation}
where $N_{x_{a}, y_{b}}(\alpha, \beta, \xi)$ is the number of
coincidences measured by the pair of detectors $x_{a},\,y_{b}$ for
$a$ and  $b$ detection apparatuses projecting photons in the above
described polarization basis. Coincident counts are measured by an
Elsag prototype of four-channel coincident circuit
\cite{elsag1,elsag2}. Single-counts and coincidences are counted
by a National Instruments \cite{disc} sixteen channels counter
plug-in PC card.

\begin{figure}[tbp]
\par
\begin{center}
\includegraphics[angle=0, width=7.5 cm, height=5 cm]{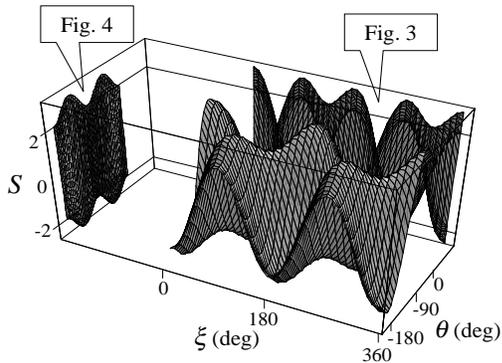}
\end{center}
\caption{ Plot of the CHSH parameter $S$ versus $\xi$ and
$\theta$. "Fig. 3" and "Fig. 4" indicates the projection of the
curve $S$ along $\theta$ and $\xi$ directions, respectively. }
\label{Figure 2}
\end{figure}

In Fig. 2 the central surface represents the theoretical value of
the CHSH parameter $S$, plotted versus $\theta$ and $\xi$. The
other two shadows labelled "Fig. 3" and "Fig. 4" are the
projections of the surface along $\theta$ and $\xi$, respectively.
These shadows highlight the theoretical bounds of $S$ for the
different values of $\theta$ and $\xi$.

Fig.s 3 and 4 show highly stable and repeatable $S$ measurement
points for many choices of $\theta$ and $\xi$. As so far
theoretically reported, the CHSH parameter satisfies Cirel'son's
inequality, and for certain values of $\theta$, there are ranges
of $S$ value not attainable by any quantum states.

\begin{figure}[tbp]
\par
\begin{center}
\includegraphics[angle=0, width=7.5 cm, height=5 cm]{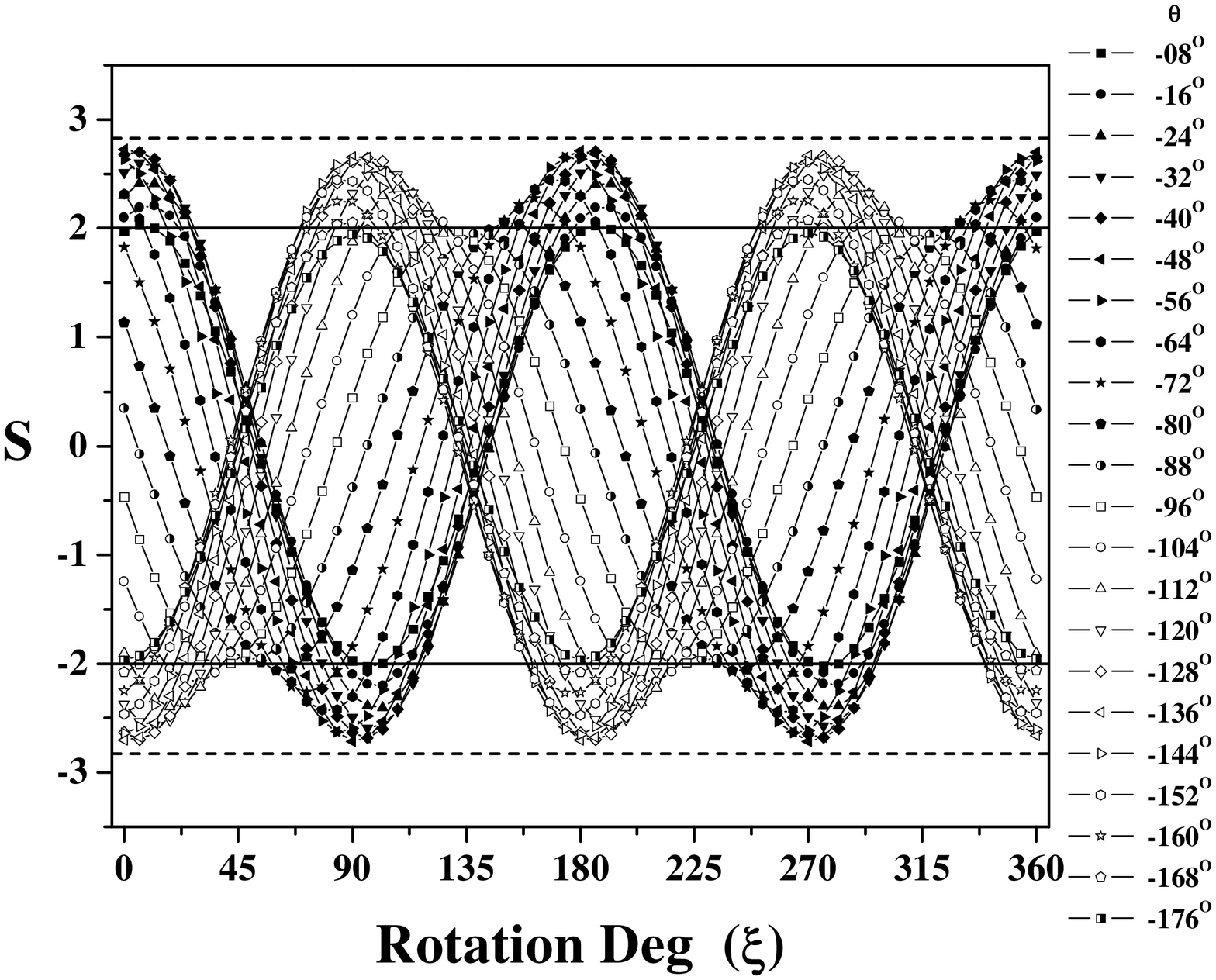}
\end{center}
\caption{ Experimental data: The CHSH parameter $S$ versus $\xi$.
Each curves corresponds to a specific choice of $\theta$. Plot
points are the measured values of $S$ for any pair of values
$\theta$ and $\xi$. } \label{Figure 3}
\end{figure}

Fig. 3 shows $S$ versus $\xi$ where the classical ($|S|=2$) and
the quantum ($|S|=2 \sqrt{2}$) limits are indicated. Each curves
corresponds to a different value of $\theta$, and these curves are
in good qualitative agreement with the theoretical predictions,
also in the case of quantum correlation bounds, as can be seen by
comparing Fig. 3 with the correspondent shadow of Fig. 2.

\begin{figure}[tbp]
\par
\begin{center}
\includegraphics[angle=0, width=7.5 cm, height=5 cm]{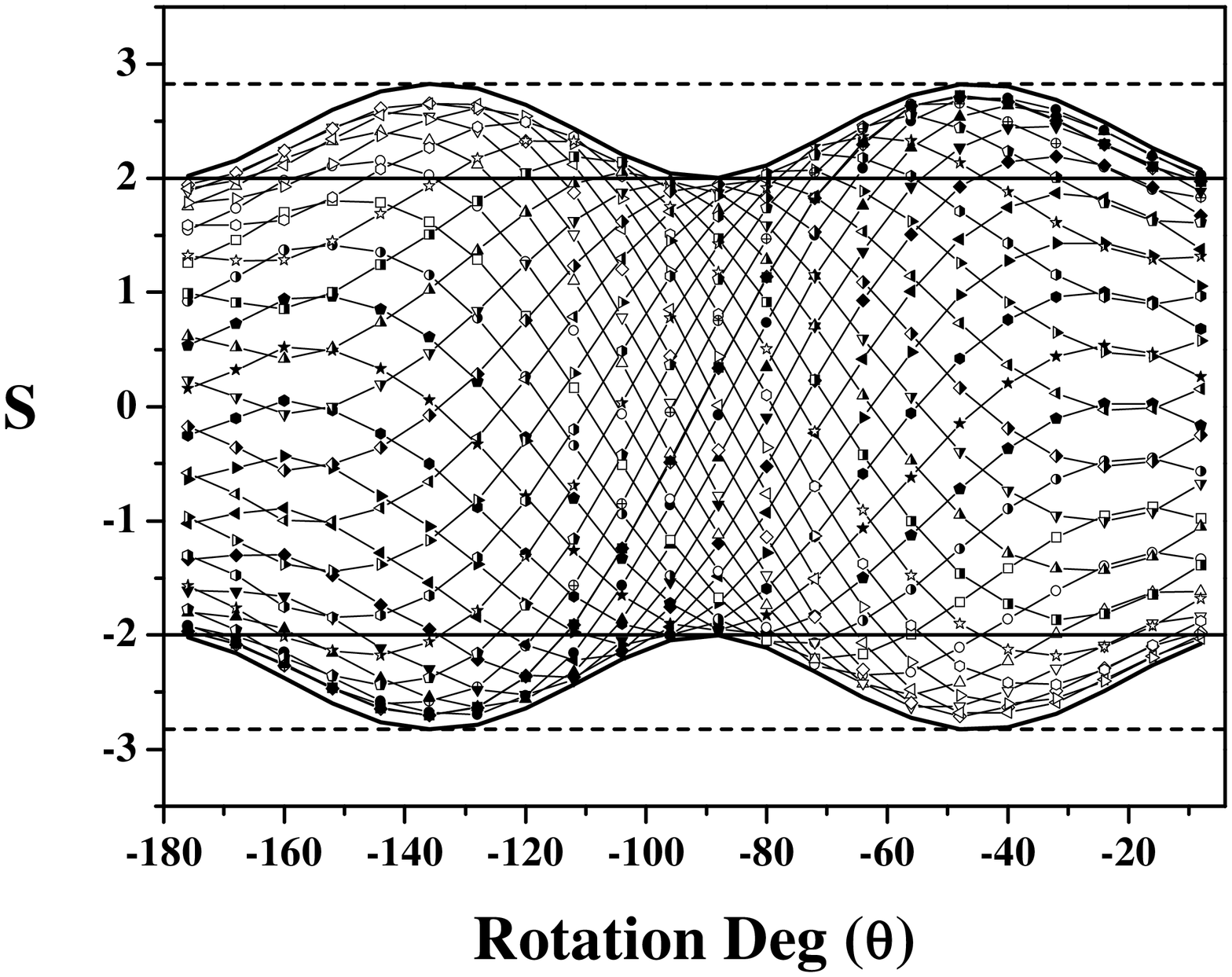}
\end{center}
\caption{ Experimental data: The CHSH parameter $S$ versus
$\theta$. Each curves corresponds to a specific choice of $\xi$.
Plot points are the measured values of $S$ for any pair of values
$\theta$ and $\xi$. } \label{Figure 4}
\end{figure}

The main thrust of the paper is reported in Fig. 4, where the
measured values of $S$ are plotted versus $\theta$ and the various
curves are associated to different values of the parameter $\xi$.
The thicker curves correspond to the theoretically predicted $S$
bounds \cite{cabellos}. There is a good qualitative and also
quantitative agreement between theoretical and experimental
bounds, even if the experimental upper (lower) bounds stands
slightly below (above) the theoretical predictions. These effects
are, as usual, imputable to noise and imperfections associated to
the polarization preservation and measurement of same setup
components, namely HWPs, PBSs, and fibers \cite{pranostro}. In
fact, the discrepancy between the theoretical and the experimental
results observed in this experiment is confirmed by an equivalent
noise level as verified during the alignment process, e.g. at
specific trivial angle settings of the polarizers.

In conclusion we performed the appealing experiment proposed in
Ref. \cite{cabellos}, to investigate the bounds of the quantum
correlation. According to quantum mechanics we did not observe any
violation of Cirel'son's inequality, with the experimental
measured bounds in agreement with the predicted ones within the
experimental known limitations; thus we can assert that our
experimental results confirm the predictions of quantum mechanics.

Despite the fact that there is no plausible theory which helps us
to design a state which violates the quantum correlation bounds,
we underline that the actual experiment is not able to search for
these violation. Quantum mechanics predicts that no quantum state
can violate these quantum bounds for any value of $\theta$. To
verify, at least partially, this last statement we are working on
the design of a new state-source to span portion of the
two-qubit-Hilbert space larger than the one identified by Eq.
(\ref{eq4}).

In our opinion the above described theory and experiment can be of
relevance in the big picture of controlling, manipulating, and
mapping quantum states, specifically in what this implies in the
emerging field of quantum technologies.

We thanks A. Cabello and M. L. Rastello for the encouragement and
for helpful suggestions. We are indebted to A. M. Colla, P.
Varisco, A. Martinoli, P. De Nicolo, S. Bruzzo, M. Genovese, I.
Ruo Berchera. This experiment was carried out in the Quantum
Optics Laboratory of Elsag S.p.A., Genova (Italy), within a
project entitled "Quantum Cryptographic Key Distribution"
co-funded by the Italian Ministry of Education, University and
Research (MIUR) - grant n. 67679/ L. 488.


\begin{thebibliography}{99}
\bibitem{disc}  Certain trade names and company products are mentioned in the text
or identified in an illustration in order to adequately specify
the experimental procedure and equipment used. In no case does
such identification imply recommendation or endorsement by the
Istituto Elettrotecnico Nazionale G. Ferraris, or by ELSAG S.p.A.,
nor does it imply that the products are necessarily the best
available for the purpose.
\end{thebibliography}
\end{document}